%% file: main.tex
\documentclass[sigconf]{acmart}

\setcopyright{acmcopyright}
\copyrightyear{2022}
\acmYear{2022}
\acmDOI{}

\acmConference[DAC '22]{Design Automation Conference}{July 10--14, 2022}{San Francisco, CA}
\acmPrice{}
\acmISBN{}

\input{mainmac}

\begin{document}

\title{Automatic oracle generation in Microsoft's Quantum Development Kit using QIR and LLVM passes}
\subtitle{Invited paper}
\author{Mathias Soeken}
\affiliation{%
  \institution{Microsoft Quantum}
  \city{Zürich}
  \country{Switzerland}
}
\email{mathias.soeken@outlook.com}
\author{Mariia Mykhailova}
\affiliation{%
  \institution{Microsoft Quantum}
  \city{Redmond}
  \state{WA}
  \country{USA}
}
\email{mamykhai@microsoft.com}

\begin{abstract}
Automatic oracle generation techniques can find optimized quantum circuits for
classical components in quantum algorithms. However, most implementations of
oracle generation techniques require that the classical component is expressed
in terms of a conventional logic representation such as logic networks, truth
tables, or decision diagrams. We implemented LLVM passes that can automatically
generate QIR functions representing classical Q\# functions into QIR code
implementing such functions quantumly. We are using state-of-the-art logic
optimization and oracle generation techniques based on XOR-AND graphs for this
purpose. This enables not only a more natural description of the quantum
algorithm on a higher level of abstraction, but also enables
technology-dependent or application-specific generation of the oracles.

\end{abstract}

\maketitle

\section{Introduction}
Implementing quantum oracles is difficult. Classical Boolean oracles are treated
as black boxes in description of algorithms (see, e.g., Hamiltonian
simulation~\cite{BCK15}, numerical gradient estimation~\cite{Jordan05}, or
amplitude amplification~\cite{Grover96,BHMT02}). While the \textit{quantum
parts} (reflection operator in Grover, QFT in QPE) of the algorithm are
described in detail, the oracle is just a placeholder with no implementation or
at best an an example for a simple classical function. In this paper, we show
how to make use of the LLVM~\cite{LA04} infrastructure to create a
QIR-based\footnote{https://qir-alliance.org} tool that can automatically
generate Q\# operations~\cite{SGT+18,HSM+20} for such classical oracles from Q\#
functions.

With the help of our approach, a developer can write a quantum program as
follows:
\begin{lstlisting}[style=Qsharp]
namespace Operations.Classical {
  internal function Majority3(a: Bool, b: Bool, c: Bool): Bool {
    return (a or b) and (a or c) and (b or c);
  }
}

namespace Operations {
  operation Majority3(
    inputs : (Qubit, Qubit, Qubit),
    output : Qubit
  ): Unit {}

  @EntryPoint()
  operation Program(): Unit {
    use (a, b, c) = (Qubit(), Qubit(), Qubit());
    use y = Qubit();

    Majority3((a, b, c), y);
  }
}
\end{lstlisting}
The program contains an internal classical function \code{Majority3} that takes
as input 3~Boolean arguments and returns a single Boolean value that is true if
and only if the majority of the input arguments is true.  The operation
\code{Majority3}, with the same name but a different namespace, is empty and
will be derived using our approach.  It can then be automatically used anywhere
in the code, e.g., in the \code{Program} operation shown in the sample.

In our approach we investigate the case in which the classical input function is
defined over tuples of Boolean input arguments and returns tuples of Boolean
output values.  Since Q\# functions are side-effect free, such functions can be
represented by combinational Boolean functions $f : \mathds{B}^n \to
\mathds{B}^m$.  We will first derive such a Boolean function from the LLVM code
generated for the Q\# function.  This function is then mapped to a quantum
circuit using state-of-the-art logic synthesis based quantum compilation
algorithms (see, e.g., \cite{MS12,MS13,Rawski15,SRWM19,MSC+19,MSM22}).  Finally,
the quantum circuit is mapped to QIR and combined with the LLVM file generated
for the rest of the Q\# program before linking.  This enables a seamless
experience, in which the user is relieved from the burden of implementing the
classical function as a quantum operation.  In addition, the automatic quantum
compilation tools employ logic optimization to reduce the implementation cost
for the oracles in terms of operation depth and qubit count.

We implemented the proposed approach in C++ and embedded it into a CMake
compilation script.  The complete implementation is publicly available as part
of the Microsoft Quantum Development Kit
samples.\footnote{https://github.com/microsoft/Quantum/tree/main/samples/qir/oracle-generator}

\section{Related work}
In~\cite{SHR18}, the authors have shown how to integrate phase oracle synthesis
and permutation synthesis into ProjectQ~\cite{SHT16} using RevKit~\cite{SFWD11}
and how to integrate automatic permutation synthesis in Q\#~\cite{SGT+18}.
In~\cite{AG20}, the authors describe a quantum programming flow that allows to
augment openQASM programs with Verilog code, which is automatically translated
into quantum circuits using tweedledum~\cite{SRHM18}. In~\cite{BBGV20}, the
authors present the quantum programming language silq, which maps common
arithmetic operations to quantum operations and supports automatic
uncomputation.

\section{Workflow}
\begin{figure}[t]
\centering
\input{figures/flow.tex}
\caption{Proposed compilation flow}
\label{fig:flow}
\end{figure}
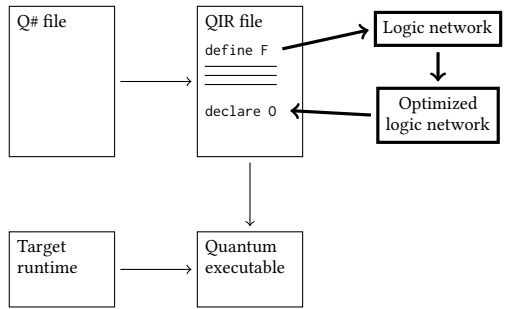
In this section, we illustrate the complete workflow of our proposed approach by
walking through the running \code{Majority3} example from the introduction.
Fig.~\ref{fig:flow} shows an overview of the overall flow.  The Q\# file is
translated into QIR using the Q\# compiler.  This file contains an LLVM function
\emph{definition} \code{F} for the classical Q\# function for which an oracle
should be generated and an LLVM function \emph{declaration} \code{O} (without
code body) for the empty operation.  A logic network is then created for
\code{F}, which is further optimized using logic synthesis techniques.  More
precisely, we will make use of XOR-AND-inverter graphs (XAGs), which are logic
networks consisting of binary XOR and binary AND gates, in which signals may be
inverted.  From the logic network, LLVM code is generated to be used as the code
body for \code{O}.  The resulting QIR file is compiled into a quantum executable
using additional information from a target runtime. All thick arrows and boxes
are contributions of our proposed flow on top of the existing QIR compilation
flow. Each subsection in the remainder describes one of the arrows in the
figure.

\subsection{Translating LLVM into a logic network}
The initial LLVM code for the \code{Majority3} function in Q\# function looks as
follows:
\begin{lstlisting}[language=llvm]
define internal i1 @Classical_Majority3(i1 %a, i1 %b, i1 %c) {
entry:
  %0 = or i1 %a, %b
  br i1 %0, label %condTrue__1, label %condContinue__1

condTrue__1:
  %1 = or i1 %a, %c
  br label %condContinue__1

condContinue__1:
  %2 = phi i1 [ %1, %condTrue__1 ], [ %0, %entry ]
  br i1 %2, label %condTrue__2, label %condContinue__2

condTrue__2:
  %3 = or i1 %b, %c
  br label %condContinue__2

condContinue__2:
  %4 = phi i1 [ %3, %condTrue__2 ], [ %2, %condContinue__1 ]
  ret i1 %4
}
\end{lstlisting}
We do not cover LLVM's syntax in detail, but describe few concepts that are
important for the remainder of the paper.  Every variable, function, and
statement is typed.  The type \code{i1} describes a 1-bit integer type, which
can encode a Boolean value.  Variable names are prefixed by a \code{\%}.  Lines
such as \code{entry:} and \code{condTrue\_\_1:} are labels and mark the
beginning of a basic block.  Each basic block contains a continuous sequence of
statements.  Statements may produce a value which is assigned to some variable.
The basic block \code{entry:} marks the first basic block of the function.
Examples for LLVM statements are \code{or} which computes the Boolean or of two
variables and stores it into a result variable, or the \code{br} which jumps to
some basic block unconditionally.

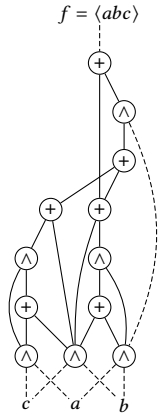
\begin{figure}[t]
\centering
\input{figures/maj3-initial.tex}
\caption{Initial logic network generated from LLVM code}
\label{fig:maj3-initial}
\end{figure}

The translation makes use of static single assignment (SSA)
$\phi$-nodes~\cite{LA04}, however, their implicitness make an automatic
translation into logic networks difficult.  We therefore make use of LLVM's
\code{reg2mem} transformation pass which introduces explicit memory instructions
to explicitly store intermediate variables.  Since our approach only parses the
transformed result to create a logic network (and not to execute it as is),
these instructions will not cause any overhead in memory access in the target
program.  The transformed function looks as follows:
\begin{lstlisting}[language=llvm,breaklines=true,postbreak=\mbox{$\hookrightarrow$\space}]
define internal i1 @Classical_Majority3(i1 %a, i1 %b, i1 %c) {
entry:
  %.reg2mem = alloca i1, align 1
  %.reg2mem1 = alloca i1, align 1
  %.reg2mem4 = alloca i1, align 1
  %.reg2mem6 = alloca i1, align 1
  %.reg2mem9 = alloca i1, align 1
  %.reg2mem11 = alloca i1, align 1
  %"reg2mem alloca point" = bitcast i32 0 to i32
  %0 = or i1 %a, %b
  store i1 %0, i1* %.reg2mem6, align 1
  %.reload8 = load i1, i1* %.reg2mem6, align 1
  br i1 %.reload8, label %condTrue__1, label %entry.condContinue__1_crit_edge

entry.condContinue__1_crit_edge:
  %.reload7 = load i1, i1* %.reg2mem6, align 1
  store i1 %.reload7, i1* %.reg2mem11, align 1
  br label %condContinue__1

condTrue__1:
  %1 = or i1 %a, %c
  store i1 %1, i1* %.reg2mem4, align 1
  %.reload5 = load i1, i1* %.reg2mem4, align 1
  store i1 %.reload5, i1* %.reg2mem11, align 1
  br label %condContinue__1

condContinue__1:
  %.reload12 = load i1, i1* %.reg2mem11, align 1
  store i1 %.reload12, i1* %.reg2mem1, align 1
  %.reload3 = load i1, i1* %.reg2mem1, align 1
  br i1 %.reload3, label %condTrue__2, label %condContinue__1.condContinue__2_crit_edge

  condContinue__1.condContinue__2_crit_edge:
  %.reload2 = load i1, i1* %.reg2mem1, align 1
  store i1 %.reload2, i1* %.reg2mem9, align 1
  br label %condContinue__2

condTrue__2:
  %2 = or i1 %b, %c
  store i1 %2, i1* %.reg2mem, align 1
  %.reload = load i1, i1* %.reg2mem, align 1
  store i1 %.reload, i1* %.reg2mem9, align 1
  br label %condContinue__2

condContinue__2:
  %.reload10 = load i1, i1* %.reg2mem9, align 1
  ret i1 %.reload10
}
\end{lstlisting}
This adds a lot of boilerplate, but most of the statements correspond to simple
operations when building the logic network.  Starting with nodes for the primary
inputs based on the function parameters \code{\%a}, \code{\%b}, and \code{\%c},
all \code{alloca} statements in the basic blocks will create temporary
variables.  These will be assigned some signal in the logic network when
\code{load} statements are encountered.  For example, in line~10 we first create
a new signal \code{\%0} by computing the OR of \code{\%a} and \code{\%b}, then
store this value in variable \code{\%.reg2mem6}, and load it into another
variable \code{\%.reload8}.  A basic block is also assigned a signal based on
the last statement in the block.  The \code{br} statement, which is the last
statement in the entry basic block, translates into a MUX operation in the logic
network.  Fig.~\ref{fig:maj3-initial} shows the resulting logic network after
the LLVM code for the function bas been completely processed.

\subsection{Optimizing the logic network}
\begin{figure}[t]
\centering
\input{figures/maj3-opt.tex}
\caption{Optimized logic network}
\label{fig:maj3-opt}
\end{figure}
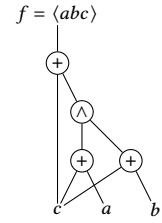

Only simple logic optimizations such as constant propagation (e.g, $1 \land x =
x$) or structural hashing~\cite{KPKG02} (never creating nodes with the same
operator and the same operands twice) are applied when creating the initial
logic network.  We can apply more advanced logic optimization techniques to the
resulting network.  It is possible to give the developer control over which cost
function to assume, or even which sequence of optimization techniques to apply.
However, a typical optimization flow would target reducing the number of AND
gates in the logic network in favor of XOR gates, since AND gates correspond to
more complicated operations both in near-term and error-corrected quantum
computing devices~\cite{AMMR13}.  This cost function relates to the
multiplicative complexity~\cite{Schnorr88} of Boolean functions.  Several
algorithms to reduce the number of AND gates in logic networks have been
proposed~\cite{BMP13,TSAM19,TSR+20,BCCM20,Soeken20}.  In our example workflow,
we employ the cut rewriting technique described in~\cite{TSAM19}.  When the
number of primary inputs does not exceed 8, we first collapse all outputs into
their truth tables and decompose them using Shannon's decomposition rule until
all internal logic nodes have~6 inputs, and use the database in~\cite{CTP19} to
map each node into its optimal XAG representation beforehand.
Fig.~\ref{fig:maj3-opt} shows the logic network after optimization.  Because the
network has a single output and not more than 6~inputs, an optimum
representation is found.

\subsection{Compiling a logic network into QIR}
There exist various techniques that describe how to map logic networks into
quantum circuits, such as~\cite{MS12,MS13,Rawski15,SRWM19,MSC+19,MSM22}.
XAG-based methods are of particular interest due to recent optimization methods
for multiplicative complexity and its close connection to resource cost of the
quantum implementation~\cite{MSC+19,MSM22}.  The mapping of an XAG into an LLVM
function based on QIR is straightforward.  One only needs function calls to
allocate and deallocate qubits, as well as calls to CNOT and Toffoli operations.
For each AND gate in the logic network, one computes the linear fanin in-place
into two existing qubits and then applies a Toffoli gate controlled on these two
qubits targeting a helper qubit.  Eventually all outputs are copied out using
CNOT gates, before all helper qubits are uncomputed by applying all operations
in reverse order.  For our running example the generated LLVM function looks as
follows (note that we shortened some function names for the quantum operations
for better readability):
\begin{lstlisting}[language=llvm,breaklines=true,postbreak=\mbox{$\hookrightarrow$\space}]
define dso_local void @Majority3__body({ %Qubit*, %Qubit*, %Qubit* }* %inputs, %Qubit* %output) {
entry:
  %0 = getelementptr inbounds { %Qubit*, %Qubit*, %Qubit* }, { %Qubit*, %Qubit*, %Qubit* }* %inputs, i32 0, i32 0
  %a = load %Qubit*, %Qubit** %0, align 8
  %1 = getelementptr inbounds { %Qubit*, %Qubit*, %Qubit* }, { %Qubit*, %Qubit*, %Qubit* }* %inputs, i32 0, i32 1
  %b = load %Qubit*, %Qubit** %1, align 8
  %2 = getelementptr inbounds { %Qubit*, %Qubit*, %Qubit* }, { %Qubit*, %Qubit*, %Qubit* }* %inputs, i32 0, i32 2
  %c = load %Qubit*, %Qubit** %2, align 8
  %qs = call %Array* @__quantum__rt__qubit_allocate_array(i64 1)
  call void @__quantum__rt__array_update_alias_count(%Array* %qs, i32 1)
  call void @CNOT(%Qubit* %c, %Qubit* %a)
  call void @CNOT(%Qubit* %c, %Qubit* %b)
  %3 = call i8* @__quantum__rt__array_get_element_ptr_1d(%Array* %qs, i64 0)
  %4 = bitcast i8* %3 to %Qubit**
  %5 = load %Qubit*, %Qubit** %4, align 8
  call void @CCNOT(%Qubit* %a, %Qubit* %b, %Qubit* %5)
  call void @CNOT(%Qubit* %c, %Qubit* %a)
  call void @CNOT(%Qubit* %c, %Qubit* %b)
  call void @CNOT(%Qubit* %c, %Qubit* %output)
  call void @CNOT(%Qubit* %5, %Qubit* %output)
  call void @CNOT(%Qubit* %c, %Qubit* %a)
  call void @CNOT(%Qubit* %c, %Qubit* %b)
  %6 = call i8* @__quantum__rt__array_get_element_ptr_1d(%Array* %qs, i64 0)
  %7 = bitcast i8* %6 to %Qubit**
  %8 = load %Qubit*, %Qubit** %7, align 8
  call void @CCNOT(%Qubit* %a, %Qubit* %b, %Qubit* %8)
  call void @CNOT(%Qubit* %c, %Qubit* %a)
  call void @CNOT(%Qubit* %c, %Qubit* %b)
  call void @__quantum__rt__qubit_release_array(%Array* %qs)
  call void @__quantum__rt__array_update_alias_count(%Array* %qs, i32 -1)
  ret void
}
\end{lstlisting}

We can provide further control to the developer by allowing to trade off the
number of helper qubits for operation count.  Reversible pebble
games~\cite{Bennett89,KSS15,MSR+19} can be used reduce the number of qubits.
This can be useful when targeting near-term devices in which the number of
qubits is highly constrained.  This approach can also support mapping algorithms
that target the operation depth rather than operation count~\cite{HS22}.

\section{Conclusion}
In this paper, we described a quantum compilation flow based on Q\#, QIR, and
LLVM, which creates quantum operations based on classical function
implementations.  The approach leverages logic networks and recent results in
logic optimization based on multiplicative complexity.  Many extensions to the
proposed flow are possible, e.g., space-efficient mapping of functions defined
over integer values, or automatic approximation-aware mapping of functions
defined over floating-point values.  Further, it is of interest to consider
mapping parameterized functions, e.g., which take a dynamic sized array as
input, since no straightforward logic network representation exists for such
cases.

\bibliographystyle{ACM-Reference-Format}
\bibliography{library,other}

\end{document}

%% file: mainmac.tex
\usepackage[utf8]{inputenc}
\usepackage[T1]{fontenc}

\usepackage{dsfont}
\usepackage{listings}
\usepackage{qsharp}
\usepackage{tikz}
\usetikzlibrary{calc}

\lstdefinelanguage{csharp}{morekeywords={var,new,typeof}}
\lstdefinestyle{csharp}{language=csharp}

\newcommand{\code}[1]{{\footnotesize\texttt{#1}}}

\tikzset{compl/.style={dash pattern=on 2pt off 1pt}}

\makeatletter
\lstdefinelanguage{llvm}{
  morecomment = [l]{;},
  morestring=[b]", 
  sensitive = true,
  classoffset=0,
  morekeywords={
    define, declare, global, constant,
    internal, external, private,
    linkonce, linkonce_odr, weak, weak_odr, appending,
    common, extern_weak,
    thread_local, dllimport, dllexport,
    hidden, protected, default,
    except, deplibs,
    volatile, fastcc, coldcc, cc, ccc,
    x86_stdcallcc, x86_fastcallcc,
    ptx_kernel, ptx_device,
    signext, zeroext, inreg, sret, nounwind, noreturn,
    nocapture, byval, nest, readnone, readonly, noalias, uwtable,
    inlinehint, noinline, alwaysinline, optsize, ssp, sspreq,
    noredzone, noimplicitfloat, naked, alignstack,
    module, asm, align, tail, to,
    addrspace, section, alias, sideeffect, c, gc,
    target, datalayout, triple,
    blockaddress
  },
  classoffset=1, keywordstyle=\color{purple},
  morekeywords={
    fadd, sub, fsub, mul, fmul,
    sdiv, udiv, fdiv, srem, urem, frem,
    and, or, xor,
    icmp, fcmp,
    eq, ne, ugt, uge, ult, ule, sgt, sge, slt, sle,
    oeq, ogt, oge, olt, ole, one, ord, ueq, ugt, uge,
    ult, ule, une, uno,
    nuw, nsw, exact, inbounds,
    phi, call, select, shl, lshr, ashr, va_arg,
    trunc, zext, sext,
    fptrunc, fpext, fptoui, fptosi, uitofp, sitofp,
    ptrtoint, inttoptr, bitcast,
    ret, br, indirectbr, switch, invoke, unwind, unreachable,
    malloc, alloca, free, load, store, getelementptr,
    extractelement, insertelement, shufflevector,
    extractvalue, insertvalue,
  },
  alsoletter={\%},
  keywordsprefix={\%},
}
\makeatother

%% file: figures/flow.tex
\begin{tikzpicture}[font=\footnotesize]
  \node[draw,minimum width=1.4cm,minimum height=2cm] (qsharp-box) {};
  \node[draw,minimum width=1.4cm,minimum height=2cm] at (2.5,0) (qir-box) {};
  \node[draw,minimum width=1.4cm,minimum height=1cm] at (2.5,-2.5) (exe-box) {};
  \node[draw,minimum width=1.4cm,minimum height=1cm] at (0,-2.5) (tgt-box) {};

  \node[draw,very thick] (init) at (5,.7) {Logic network};
  \node[draw,align=center,very thick] (opt) at (5,-.45) {Optimized \\ logic network};

  \node[anchor=north west] at (qsharp-box.north west) {Q\# file};
  \node[anchor=north west] at (qir-box.north west) {QIR file};
  \node[anchor=north west,align=left] at (exe-box.north west) {Quantum \\ executable};
  \node[anchor=north west,align=left] at (tgt-box.north west) {Target \\ runtime};

  \coordinate (cf) at ($(qir-box.north west)!0.2!(qir-box.south west)$);
  \node[font=\scriptsize\ttfamily,anchor=north west] (func) at (cf) {define F};
  \draw[shorten <=3pt,shorten >=10pt] ($(qir-box.north west)!0.4!(qir-box.south west)$) -- ($(qir-box.north east)!0.4!(qir-box.south east)$);
  \draw[shorten <=3pt,shorten >=10pt] ($(qir-box.north west)!0.46!(qir-box.south west)$) -- ($(qir-box.north east)!0.46!(qir-box.south east)$);
  \draw[shorten <=3pt,shorten >=10pt] ($(qir-box.north west)!0.52!(qir-box.south west)$) -- ($(qir-box.north east)!0.52!(qir-box.south east)$);
  \coordinate (co) at ($(qir-box.north west)!0.6!(qir-box.south west)$);
  \node[font=\scriptsize\ttfamily,anchor=north west] (op) at (co) {declare O};

  \begin{scope}[shorten <=2pt,shorten >=2pt]
    \draw[->] (qsharp-box.east) -- (qir-box.west);
    \draw[->,very thick] (func.east) -- (init.west);
    \draw[->,very thick] (init.south) -- (opt.north);
    \draw[->,very thick] (opt.west) -- (op.east);
    \draw[->] (qir-box.south) -- (exe-box.north);
    \draw[->] (tgt-box.east) -- (exe-box.west);
  \end{scope}
\end{tikzpicture}

%% file: figures/maj3-initial.tex
\begin{tikzpicture}[x=.65cm,y=0.65cm,font=\footnotesize]
  \begin{scope}[every node/.style={inner sep=.5pt}]
    \node at (0,0) (c) {$c$};
    \node at (1,0) (a) {$a$};
    \node at (2,0) (b) {$b$};
    \node at (1.5,8) (y) {$f = \langle abc\rangle$};
  \end{scope}

  \begin{scope}[every node/.style={draw,circle,inner sep=0pt,minimum size=8.5pt}]
    \node at (0,1) (n6) {$\land$};
    \node at (1,1) (n7) {$\land$};
    \node at (2,1) (n5) {$\land$};
    \node at (0,2) (n8) {$+$};
    \node at (1.5,2) (n11) {$+$};
    \node at (0,3) (n9) {$\land$};
    \node at (1.5,3) (n12) {$\land$};
    \node at (0.5,4) (n10) {$+$};
    \node at (1.5,4) (n13) {$+$};
    \node at (2,5) (n14) {$+$};
    \node at (2,6) (n15) {$\land$};
    \node at (1.5,7) (n16) {$+$};
  \end{scope}

  \draw[compl] (n6) -- (c) (n6) -- (a);
  \draw[compl] (n7) -- (c) (n7) -- (b);
  \draw[compl] (n5) -- (a) (n5) -- (b);
  \draw (n8) -- (n6) (n8) -- (n7);
  \draw (n11) -- (n5) (n11) -- (n7);
  \draw (n9) -- (n8) (n9) to[bend right=30] (n6);
  \draw (n12) -- (n11) (n12) to[bend left=20] (n5);
  \draw (n10) -- (n7) (n10) -- (n9);
  \draw (n13) to[bend right=10] (n7) (n13) -- (n12);
  \draw (n14) -- (n13) (n14) -- (n10);
  \draw (n15) -- (n14);
  \draw[compl] (n15) to[bend left=25] (n5);
  \draw (n16) -- (n13) (n16) -- (n15);
  \draw[compl] (y) -- (n16);
\end{tikzpicture}

%% file: figures/maj3-opt.tex
\begin{tikzpicture}[x=.65cm,y=0.65cm,font=\footnotesize]
  \begin{scope}[every node/.style={inner sep=.5pt}]
    \node at (0,0) (c) {$c$};
    \node at (1,0) (a) {$a$};
    \node at (2,0) (b) {$b$};
    \node at (0,4) (y) {$f = \langle abc\rangle$};
  \end{scope}

  \begin{scope}[every node/.style={draw,circle,inner sep=0pt,minimum size=8.5pt}]
    \node at (0.5,1) (n5) {$+$};
    \node at (1.5,1) (n6) {$+$};
    \node at (0.5,2) (n7) {$\land$};
    \node at (0,3) (n8) {$+$};
  \end{scope}

  \draw (n5) -- (c) (n5) -- (a);
  \draw (n6) -- (c) (n6) -- (b);
  \draw (n7) -- (n5) (n7) -- (n6);
  \draw (n8) -- (n7) (n8) -- (c);
  \draw (n8) -- (y);
\end{tikzpicture}